\documentclass[10pt]{article}

\usepackage[margin=0.9in]{geometry}
\usepackage[T1]{fontenc}
\usepackage[utf8]{inputenc}
\usepackage{amsmath,amssymb,amsthm,mathtools,bm}
\usepackage{booktabs,array}
\usepackage{graphicx}
\usepackage[hidelinks]{hyperref}
\hypersetup{
  pdftitle={QUBO Modeling of Module Learning With Errors: Stability and Scaling in Post-Quantum Cryptography},
  pdfauthor={Ruturaj Khamitkar, Durga Pritam Suggisetti, Soujanya Chatti, Varsha S. Sambhaje, Durga Dasari}
}

\title{QUBO Modeling of Module Learning With Errors: Stability and Scaling in Post-Quantum Cryptography}

\author{%
\normalsize
\begin{tabular}{c}
Ruturaj Khamitkar$^{1}$, Durga Pritam Suggisetti$^{2,3}$,\\
Soujanya Chatti$^{3}$, Varsha  Sambhaje$^{4}$$\thanks{Corresponding authors}$,\\
and Durga Dasari$^{5*}$
\\[0.8em]
\small $^{1}$D Y Patil International University, Akurdi, Maharashtra 411044, India\\
\small $^{2}$Birla Institute of Technology and Science Pilani, Dubai, UAE\\
\small $^{3}$QbitForce Quantum Pvt. Ltd., Vijayawada, AP 520012, India\\
\small $^{4}$Department of Computer Science and Engineering, SRM University-AP, Amaravati 522240, India\\
\small $^{5}$3. Physikalisches Institut, University of Stuttgart, Stuttgart 70569, Germany\\[0.4em]
\end{tabular}%
}

\date{}

\begin{document}
\maketitle

\begin{abstract}
Lattice-based post-quantum cryptography relies on the hardness of the Learning With Errors (LWE) and Module Learning With Errors (MLWE) problems. This work introduces a constructive framework for encoding small MLWE instances as Quadratic Unconstrained Binary Optimization (QUBO) models suitable for quantum annealing. The formulation jointly represents secret coefficients and explicit error variables within a unified binary optimization structure, enabling their simultaneous recovery from the ground-state solution. Beyond the encoding, we develop a stability analysis of the resulting optimization landscape under additive perturbations. We show that the admissible noise region forms a convex polytope defined by competing candidate secrets, and establish an equivalent characterization in terms of the QUBO energy gap between the optimal and second-best solutions. Numerical experiments on low-dimensional benchmark instances using exact simulation demonstrate correct recovery of both secret and discretized error vectors, and confirm consistency between geometric stability regions and energy-gap behavior. We further quantify the scaling of logical variables and embedding overhead with increasing MLWE dimensions to assess feasibility on quantum annealing architectures. The results establish a systematic connection between MLWE problems and quantum optimization while providing a framework for analyzing robustness properties of QUBO formulations. Although current quantum annealing hardware remains insufficient for cryptographically relevant parameters, the proposed methodology offers a structured basis for studying lattice-based problems in quantum optimization settings without implying a practical threat to standardized post-quantum schemes.
\end{abstract}

\noindent\textbf{Keywords:} quantum annealing, QUBO, Module Learning With Errors, lattice-based cryptography, post-quantum cryptography.

\section{Introduction}
Quantum annealing (QA) has emerged as an important optimization paradigm for problems that can be encoded as energy minimization over binary variables. Rather than executing a circuit model of computation, QA maps a cost function to an effective Hamiltonian and seeks low-energy configurations through an annealing process~\cite{Kadowaki1998,Albash2018}. This perspective has made QA attractive in application domains such as scheduling, routing, materials design, and structured combinatorial search~\cite{Venturelli2015,PerdomoOrtiz2012,Ding2024}. At the same time, the ability to express a problem as a quadratic binary objective, often in QUBO form, has become increasingly valuable beyond physics, because it provides a common language shared by quantum annealers, Ising machines, and a variety of classical heuristics.

In parallel, post-quantum cryptography (PQC) has become a central research area because widely deployed public-key systems such as RSA and elliptic-curve cryptography are vulnerable to Shor's algorithm in a sufficiently powerful quantum setting~\cite{Shor1994}. Among the main PQC families, lattice-based constructions are particularly prominent because they combine strong worst-case-to-average-case hardness evidence with efficient implementations. LWE, introduced by Regev, is one of the foundational assumptions in this area~\cite{Regev2005,Regev2009}. In its basic form, one is given noisy linear relations of the form
\begin{equation}
  b_i = \langle a_i, s \rangle + e_i \pmod q,
\end{equation}
where $s$ is the hidden secret, $a_i$ are public samples, and $e_i$ is a small error term. Structured variants such as Ring-LWE and Module-LWE (MLWE) retain the same hardness philosophy while providing the algebraic efficiency needed in practical cryptosystems such as CRYSTALS-Kyber~\cite{Lyubashevsky2010,Peikert2016,Bos2018}.

The cryptanalytic landscape of LWE and MLWE is already rich. On the classical side, the dominant approaches rely on lattice reduction, especially LLL- and BKZ-type methods, along with dual/primal attack analyses and combinatorial techniques such as BKW~\cite{Lenstra1982,Schnorr2001,Albrecht2015,Blum2003}. These methods have steadily improved concrete security estimation, but they still face severe scaling barriers for standardized parameter sets. On the quantum side, there is no known direct analogue of Shor's algorithm for general lattice problems. Instead, the literature has investigated more modest quantum advantages through quantum-enhanced search, quantum walks, and hybrid subroutines that may accelerate specific components without fundamentally collapsing the hardness assumption~\cite{Laarhoven2015,Childs2003}.

More recently, attention has turned to optimization-based formulations of lattice problems. If an LWE- or MLWE-related objective can be expressed in QUBO form,
\begin{equation*}
  \min_{x\in\{0,1\}^n} x^\top \mathcal{Q}x
\end{equation*}
then it becomes directly accessible not only to annealing hardware but also to a broader optimization ecosystem built around Ising and binary quadratic models~\cite{Lucas2014}. This has motivated a growing body of work on QUBO or Ising encodings for cryptanalytic and number-theoretic tasks, including direct annealing-style formulations, hybrid quantum--classical pipelines, coherent Ising machine approaches, and mixed-integer/QUBO perspectives on LWE-like systems~\cite{Joseph2021,Zheng2024,Jiang2026,Qayyum2025}. However, several limitations remain common: many formulations focus primarily on search or decision outputs rather than explicitly modeling both secret and error variables, and few provide a detailed stability analysis showing how perturbations affect the recovered optimum.

This paper addresses that gap by developing a constructive MLWE-to-QUBO framework in which the secret coefficients and explicit error variables are encoded jointly within a single binary optimization model. The purpose of the construction is not to claim a practical attack on deployed post-quantum schemes. Rather, it is to provide a transparent and analyzable optimization representation for small benchmark instances, and to use that representation to study the geometry and robustness of the resulting energy landscape. In particular, we examine how the optimum behaves in both noiseless and noisy settings, characterize the admissible perturbation region through a convex stability polytope, and connect that geometric picture to the QUBO energy gap between the best and second-best solutions.

The contribution should therefore be understood as methodological. We show that small MLWE instances can be encoded cleanly into QUBO form, that the same model supports simultaneous recovery of discretized secrets and noise variables, and that the associated robustness can be interpreted from both geometric and energetic viewpoints. We also quantify the scaling of logical variables and explain why dense connectivity and embedding overhead quickly place cryptographically relevant instances beyond current quantum annealing hardware. Taken together, these considerations position the present work as a bridge between lattice-based cryptographic modeling and hardware-aware binary optimization, rather than as a practical attack on deployed post-quantum schemes.

\section{Encoding of MLWE as QUBO}
\begin{table}[t]
  \centering
  \caption{Notation used throughout the MLWE--QUBO formulation.}
  \label{tab:notation}
  \begin{tabular}{@{}ll@{}}
    \toprule
    Symbol & Meaning \\
    \midrule
    $A$ & Public MLWE matrix over $\mathbb{Z}_q$ \\
    $s$ & Secret vector to be recovered \\
    $e$ & Error (noise) vector \\
    $b$ & Observation vector, $b = As + e \pmod{q}$ \\
    $q$ & Modulus of the MLWE instance \\
    $N_{\text{logical}}$ & Number of logical binary variables in the QUBO representation \\
    $\mathcal{Q}$ & Interaction matrix of the QUBO model \\
    $x$ & Binary variable vector in the QUBO encoding \\
    \bottomrule
  \end{tabular}
\end{table}

\subsection{Problem Formulation}
Lattice-based cryptographic schemes rely on the hardness of structured problems over integer lattices. In MLWE, a public matrix $A \in \mathbb{Z}_q^{m\times n}$, a secret vector $s \in \mathbb{Z}_q^n$, and a noise vector $e \in \mathbb{Z}_q^m$ generate an observation vector
\begin{equation}
  b = As + e \pmod{q}.
\end{equation}
The goal of MLWE recovery is to estimate $s$ given $(A,b)$ in the presence of noise.

To obtain an optimization formulation, the modular constraint is lifted to the integers and expressed as a residual minimization problem. For a candidate solution $s'$, we define the residual
\begin{equation*}
  r(s') = As' - b.
\end{equation*}
This leads to the energy function
\begin{equation}
  E(s') = \|As' - b\|_2^2.
\end{equation}
To enable a QUBO representation, both the secret variables and the error components are explicitly encoded into binary variables through a finite-bit representation of the integer domain. Let $x \in \{0,1\}^{N_{\text{logical}}}$ denote the resulting binary vector encoding both $s$ and $e$. Under this encoding, the energy function can be rewritten as a quadratic polynomial in binary variables,
\begin{equation*}
  E(x) = x^\top \mathcal{Q}x + c^\top x,
\end{equation*}
which defines the corresponding QUBO formulation.

\subsection{Zero-Noise Case ($e=0$)}
When $e=0$, the MLWE instance reduces to a deterministic quadratic optimization problem over the secret variables alone. Each coefficient $s_i\in\mathbb{Z}_q$ is represented in binary form as
\begin{equation*}
  s_i = \sum_{k=0}^{\lceil \log_2 q\rceil-1} 2^k x_{ik}, \qquad x_{ik}\in\{0,1\}.
\end{equation*}
Substituting this representation into the energy function $E(s') = \|As' - b\|_2^2$ yields a quadratic polynomial in binary variables. This results in a standard QUBO formulation of the form
\begin{equation*}
  \min_{x\in\{0,1\}^N} x^\top \mathcal{Q}x + c^\top x
\end{equation*}
where $N = n\lceil \log_2 q\rceil$. This formulation corresponds to a pure optimization landscape over the secret space without noise-induced coupling effects.

\subsection{Nonzero-Noise Case ($e\neq 0$)}
For general MLWE instances, we consider a joint optimization over both secret and noise variables. The corresponding objective is
\begin{equation*}
  \min_{s,e} \; \|As + e - b\|_2^2 + \lambda \|e\|_2^2.
\end{equation*}
where $\lambda>0$ controls the relative penalty on noise magnitude and ensures bounded solutions. For the toy benchmark studied here we use $\lambda=0.1$ as a mild regularization choice; this value is intended only for the illustrative low-dimensional example and is not claimed to be universal.

Unlike formulations that eliminate the noise term, this representation explicitly models $e$ as an optimization variable, enabling direct analysis of its interaction with the secret. Both $s$ and $e$ are encoded in binary form as
\begin{equation*}
  s_i = \sum_k 2^k x_{ik}, \qquad e_j = \sum_\ell 2^\ell y_{j\ell}.
\end{equation*}
Defining the combined binary vector $z=[x;y]$, the resulting QUBO can be written as
\begin{equation*}
  \min_{z\in\{0,1\}^{N_{\text{logical}}}} z^\top \mathcal{Q}z + c^\top z
\end{equation*}
The interaction matrix of the QUBO model admits the block structure
\begin{equation*}
  \mathcal{Q} = \begin{bmatrix} \mathcal{Q}_{ss} & \mathcal{Q}_{se} \\ \mathcal{Q}_{es} & \mathcal{Q}_{ee} \end{bmatrix}
\end{equation*}
where $\mathcal{Q}_{ss}$ encodes interactions among secret variables, $\mathcal{Q}_{ee}$ encodes noise regularization terms, and $\mathcal{Q}_{se}$ captures the coupling between secret and noise components, which governs the coupled energy landscape and stability behavior.

Unlike prior formulations based on penalty-free coherent Ising machine embeddings~\cite{Jiang2026} or hybrid NISQ approaches~\cite{Zheng2024}, the present work explicitly retains both secret and noise degrees of freedom within a unified QUBO representation, enabling direct analysis of solution stability under perturbations.

\section{Implementation of a Small Instance}
\subsection{Case (i): $e=0$}
We first consider a noise-free MLWE instance as a baseline validation of the proposed encoding. For this toy benchmark we take $q=4$, consistent with the 2-bit encoding used for each secret coefficient:
\begin{equation*}
  A=
  \begin{bmatrix}
  1 & 2 & 0 \\
  0 & 1 & 1 \\
  2 & 0 & 1
  \end{bmatrix}, \qquad
  s=
  \begin{bmatrix}
  1 \\ 0 \\ 1
  \end{bmatrix}, \qquad
  b=As.
\end{equation*}
Each secret coefficient is encoded using a 2-bit binary representation:
\begin{equation*}
  s_i = x_{2i-1} + 2x_{2i}, \qquad x_i\in\{0,1\}.
\end{equation*}
The resulting objective reduces to
\begin{equation*}
  \min_x \|As-b\|^2 = x^\top \mathcal{Q}x + c^\top x,
\end{equation*}
which is solved exactly by the QUBO formulation. The recovered solution matches the ground-truth secret:
\begin{equation*}
  s=(1,0,1).
\end{equation*}

\begin{figure}[t]
  \centering
  \includegraphics[width=\linewidth]{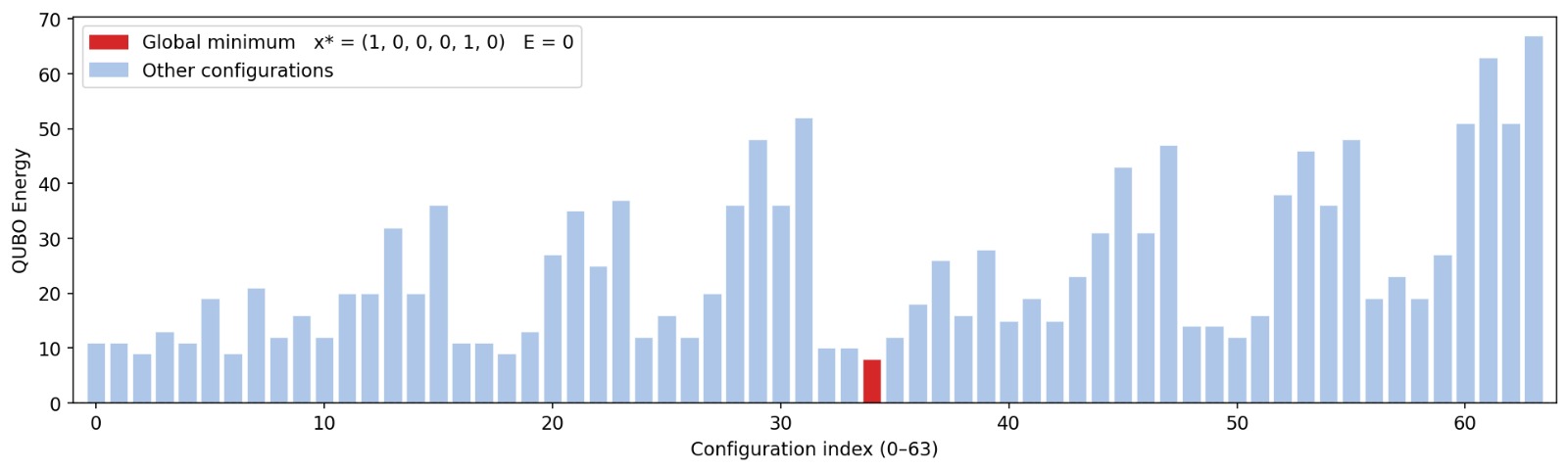}
  \caption{Case (i), $e=0$. Exhaustive evaluation of all $2^6$ configurations of the corrected QUBO energy for the toy benchmark. The minimum-energy configuration corresponds to $x^\ast=[1,0,0,0,1,0]$, decoding to $s=(1,0,1)$ with zero residual energy.}
  \label{fig:case1}
\end{figure}

\subsection{Case (ii): $e\neq 0$}
We now consider the general noisy MLWE setting for the same toy modulus $q=4$:
\begin{equation*}
  b=As+e,
\end{equation*}
with $A\in \mathbb{Z}_q^{m\times n}$, $s\in\mathbb{Z}_q^n$, and $e\in\mathbb{R}^m$ representing bounded perturbations.

We formulate joint recovery of $s$ and $e$ via the optimization problem
\begin{equation*}
  \min_{s,e}\; \|As+e-b\|_2^2 + \lambda \|e\|_2^2.
\end{equation*}
Each variable is encoded in binary form:
\begin{equation*}
  s_i = \sum_k 2^k x_{ik}, \qquad e_j = \sum_\ell 2^\ell y_{j\ell},
\end{equation*}
with $x_{ik},y_{j\ell}\in\{0,1\}$. Let $z=[x;y]$ denote the combined binary vector. Substituting the encodings into the objective yields a quadratic function
\begin{equation*}
  E(z)=z^\top \mathcal{Q}z + c^\top z,
\end{equation*}
where $\mathcal{Q}$ admits the block structure
\begin{equation*}
  \mathcal{Q}=
  \begin{bmatrix}
    \mathcal{Q}_{ss} & \mathcal{Q}_{se} \\
    \mathcal{Q}_{es} & \mathcal{Q}_{ee}
  \end{bmatrix}.
\end{equation*}
The block $\mathcal{Q}_{ss}$ corresponds to the secret variables, $\mathcal{Q}_{ee}$ arises from noise regularization, and $\mathcal{Q}_{se}$ captures coupling induced by the residual term $As+e-b$.

For a given matrix $A$ and secret vector $s$, introducing the noise vector $e$ increases the dimensionality of the optimization problem. In this worked example we use a 4-bit shifted discretization for each noise component, so the QUBO variable space expands from $n\lceil \log_2 q\rceil$ variables to $n\lceil \log_2 q\rceil + 4m$ variables. Consequently, the interaction matrix $\mathcal{Q}$ scales from a $6\times 6$ structure (in the noise-free case) to an $18\times 18$ structure for the joint estimation problem in this example.

Solving the resulting QUBO yields simultaneous recovery of both the secret and noise components,
\begin{equation*}
  s=(1,0,1), \qquad e=(0.2,-0.2,0.3),
\end{equation*}
up to discretization induced by the binary encoding of $e$. This formulation explicitly embeds noise as an optimization variable, enabling analysis of solution stability under perturbations and yielding a richer energy landscape than noise-free MLWE encodings.

\section{Numerical Simulations}
This section reports numerical results obtained from direct evaluation of the MLWE-to-QUBO formulation described in the previous sections. All figures reproduce the behavior of the noiseless and noisy cases, including stability characteristics and scaling trends. The underlying objective functions and instance parameters are not modified.

The numerical evaluation begins with the noiseless MLWE instance in which $e=0$. All $2^6$ binary configurations are exhaustively enumerated, allowing complete characterization of the QUBO energy landscape. Figure~\ref{fig:case1} illustrates that the global minimum corresponds to the correct secret vector, with a clear separation between the optimal configuration and competing states. This distinct minimum confirms that the proposed encoding accurately represents the underlying optimization problem in the absence of noise.

\begin{figure}[t]
  \centering
  \includegraphics[width=\linewidth]{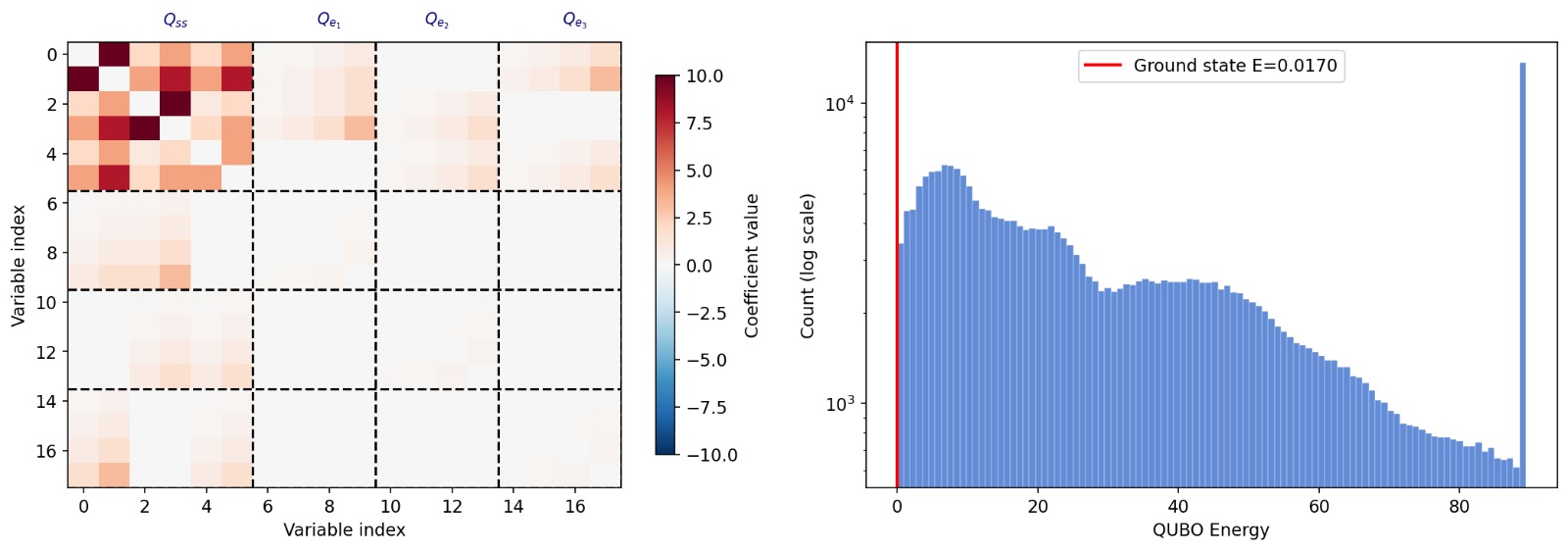}
  \caption{Case (ii), $e\neq 0$. Left: $18\times 18$ interaction matrix $\mathcal{Q}$ showing block structure between secret and noise variables. Right: exhaustive energy distribution over $2^{18}$ configurations, showing a unique global minimum.}
  \label{fig:case2}
\end{figure}

The analysis is then extended to the noisy setting by introducing the perturbation vector $e=(0.2,-0.2,0.3)$ and constructing the joint secret--noise QUBO containing 18 binary variables. Figure~\ref{fig:case2} shows that the corresponding interaction matrix $\mathcal{Q}$ naturally separates into secret, noise, and coupling blocks arising from the residual term. Exhaustive evaluation of all configurations reveals a unique global minimum corresponding to the correct joint assignment, demonstrating that the formulation successfully incorporates explicit noise variables while preserving an identifiable optimum.

\section{Illustrative Stability Geometry Under Noise}
The recovery of the correct secret in an MLWE instance depends not only on the optimization algorithm but also on the magnitude and direction of the perturbation introduced by the error vector. In the full MLWE-QUBO model of this paper, the relevant objective depends on the transformed residual $As-b$, so exact decision boundaries are matrix dependent. The analysis in this section is therefore presented explicitly as an illustrative geometric surrogate: it visualizes robustness in secret space through a nearest-neighbor model, rather than claiming to be a complete derivation of the exact $A$-dependent MLWE-QUBO stability condition. Within that scope, it provides an interpretable picture of how perturbations can alter the preferred decoded solution.

Let $s\in\mathbb{R}^n$ denote the true secret and let the observed vector be perturbed by a noise vector $e$. In the surrogate model, the stability region is defined as the set of all perturbations for which $s$ remains the minimizer of the corresponding nearest-neighbor problem,
\begin{equation*}
  \min_{s'\in S} \|s'-(s+e)\|^2,
\end{equation*}
where $S$ denotes the set of all candidate secrets. The true secret is therefore recovered whenever
\begin{equation*}
  \|e\|^2 < \|e+(s-s')\|^2, \qquad \forall s'\neq s.
\end{equation*}
Expanding the quadratic terms eliminates the common norm of the perturbation and yields a linear inequality for every competing secret,
\begin{equation*}
  2(s-s')^\top e + \|s-s'\|^2 > 0.
\end{equation*}
Each competing candidate therefore defines a half-space in the noise domain. The admissible perturbations are obtained by intersecting all such half-spaces, producing a convex polytope within which the decoded secret remains unchanged. Consequently, robustness against noise is determined by the geometry of the candidate-secret constellation rather than by the magnitude of the perturbation alone.

\begin{figure}[t]
  \centering
  \includegraphics[width=0.72\linewidth]{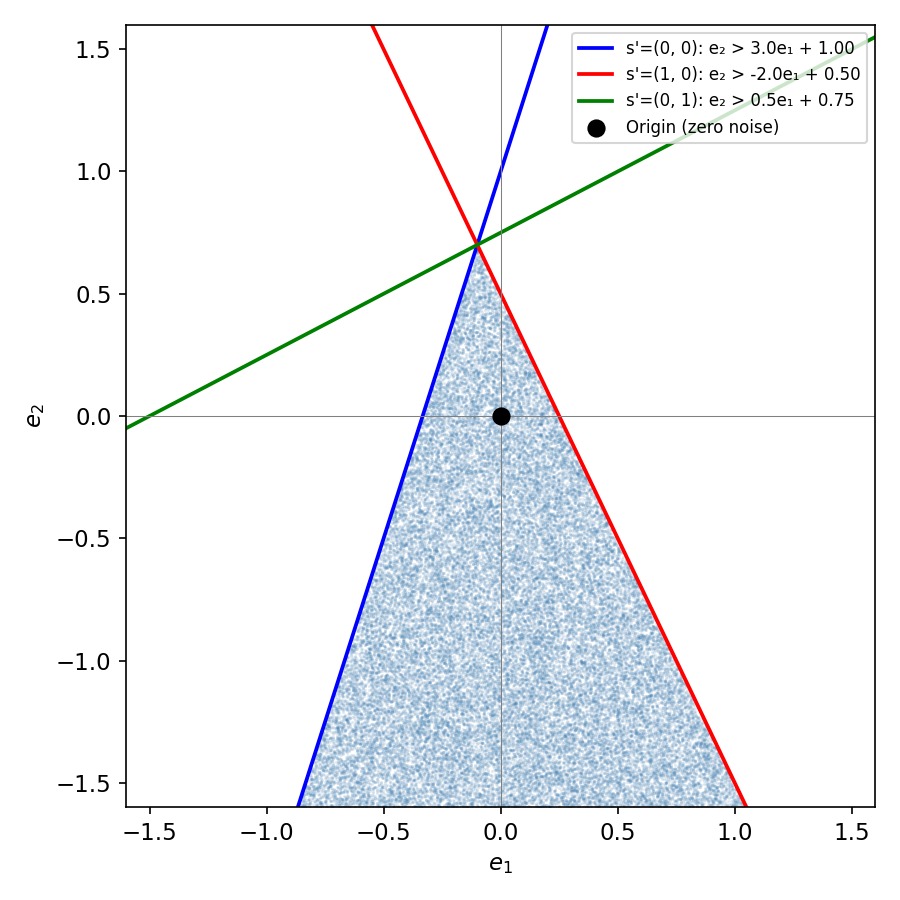}
  \caption{Two-dimensional stability region for the MLWE instance with $s=(0.6,-0.2)$. Each boundary line corresponds to a competing candidate secret and defines a half-space constraint in the noise variables $(e_1,e_2)$. The intersection of these half-spaces forms a convex polytope representing all perturbations that preserve correct recovery of the true secret.}
  \label{fig:stability-fig3}
\end{figure}

To illustrate the construction, consider the two-dimensional example
\begin{equation*}
  s=(0.6,-0.2), \qquad S=\{(0,0),(1,0),(0,1)\}.
\end{equation*}
For this example the three competing candidates generate the half-plane constraints
\begin{align*}
  s'=(0,0): &\qquad e_2 < 3e_1 + 1, \\
  s'=(1,0): &\qquad e_2 < -2e_1 + 0.5, \\
  s'=(0,1): &\qquad e_2 < 0.5e_1 + 0.75.
\end{align*}
The intersection of these inequalities forms the stability region shown in Figure~\ref{fig:stability-fig3}. Each boundary corresponds to one competing secret, while the shaded polytope represents the complete set of admissible perturbations that preserve the decoded solution. As long as the noise vector remains inside this region, the optimization problem continues to identify the correct secret. Crossing any boundary causes another candidate to become energetically preferable, resulting in a decoding error.

\begin{figure}[t]
  \centering
  \includegraphics[width=\linewidth]{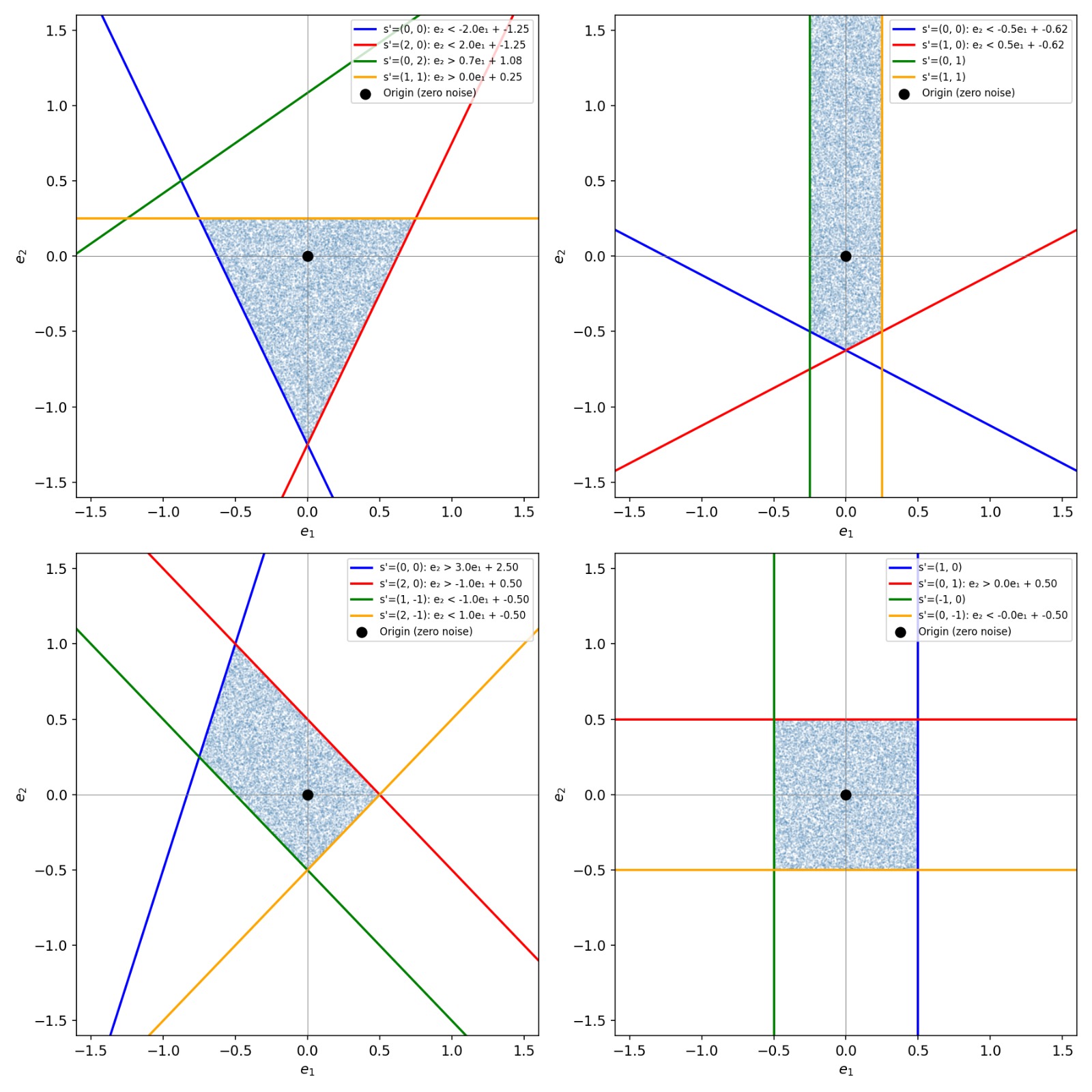}
  \caption{Stability polytopes for four different true secrets constructed using the same half-space formulation. The shaded regions represent the intersection of linear constraints induced by competing candidate secrets, forming convex polytopes that characterize recovery stability. Larger or more symmetric polytopes correspond to increased tolerance to perturbations before a change in the optimal decoded solution occurs.}
  \label{fig:stability-grid}
\end{figure}

The half-plane formulation naturally generalizes to arbitrary candidate sets. Figure~\ref{fig:stability-grid} presents stability polytopes for four different true secrets generated using the same construction. Although the mathematical formulation is identical in every case, the admissible regions differ significantly in both shape and size because they are determined by the relative positions of the true secret and its competitors. Larger and more symmetric polytopes indicate that a wider range of perturbations can be tolerated before the optimal solution changes, whereas narrow or irregular regions correspond to reduced robustness.

While the geometric construction provides an intuitive description of robustness, the same surrogate phenomenon can also be interpreted energetically through the separation between the best and second-best configurations. Let $E_1$ and $E_2$ denote the lowest and second-lowest QUBO energies, respectively. The quantity
\begin{equation*}
  \Delta E = E_2 - E_1
\end{equation*}
defines the energy gap separating the optimal solution from its nearest competitor. A positive energy gap implies that the correct secret is uniquely distinguishable, whereas $\Delta E = 0$ indicates that two competing configurations become equally favorable, marking the onset of recovery failure.

\begin{figure}[t]
  \centering
  \includegraphics[width=0.85\linewidth]{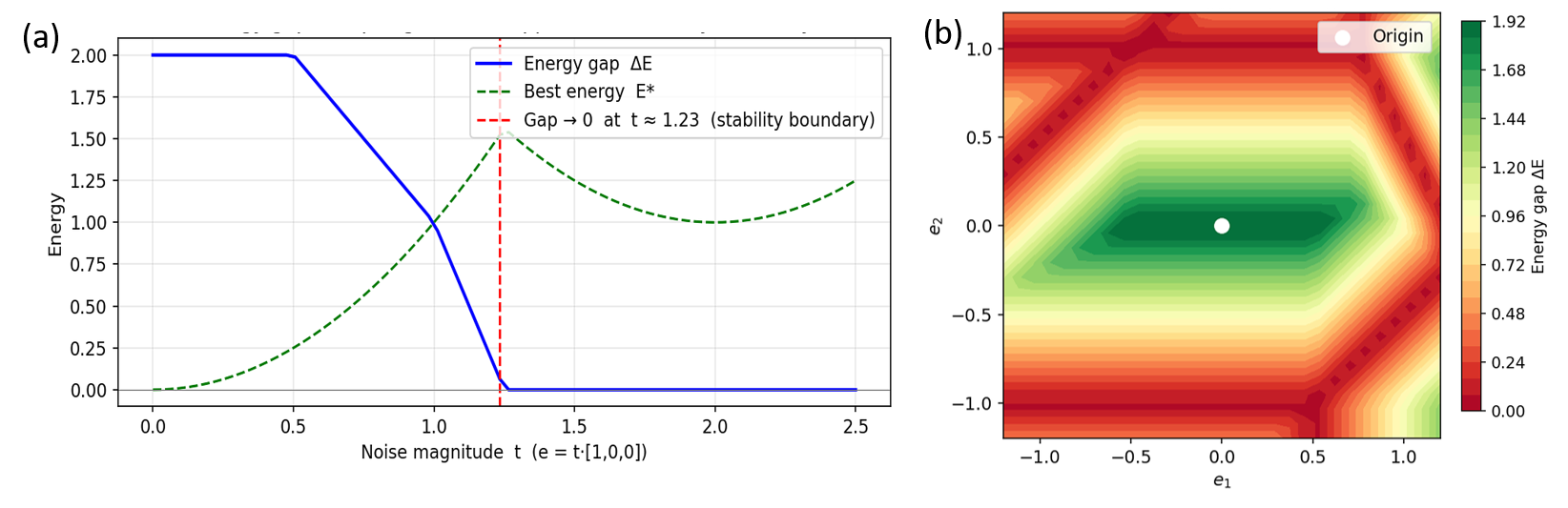}
  \caption{(a) Energy-gap evolution under increasing noise magnitude $t$, showing gap closure at a critical threshold and loss of correct recovery. (b) Energy-gap heatmap in the $(e_1,e_2)$ plane with $e_3=0$, where the zero-gap contour is consistent with the geometric stability region.}
  \label{fig:energy-gap}
\end{figure}

Figure~\ref{fig:energy-gap}(left) illustrates this behavior by increasing the perturbation magnitude along a fixed direction in noise space. Initially, the energy gap remains positive and the recovered secret is stable. As the perturbation increases, the gap gradually decreases until it vanishes at a critical threshold, beyond which the original secret is no longer the unique minimizer of the QUBO objective.

A more complete view is obtained by evaluating the surrogate energy gap over a two-dimensional perturbation grid. Figure~\ref{fig:energy-gap}(right) shows that regions with a large positive gap coincide with the interior of the geometric stability region derived from the half-plane construction. The contour where the gap vanishes reproduces the same boundary, demonstrating that the geometric and energetic viewpoints are equivalent within this simplified model.

The proposed surrogate stability framework therefore provides two complementary interpretations of robustness. The geometric perspective characterizes an admissible perturbation region as a convex polytope defined by competing candidate secrets, whereas the energetic perspective measures robustness through the separation between the optimal solution and its nearest competitor in the simplified landscape. Together, these viewpoints explain qualitatively how recovery can fail as the perturbation approaches a decision boundary, while leaving a full matrix-aware MLWE-QUBO stability analysis to future work.

\section{Generalization, Scaling, and Hardware Limits}
The QUBO formulation presented for small MLWE instances generalizes naturally to larger dimensions, but its scalability is fundamentally constrained by the growth in binary variables and quadratic couplings. For a general instance with
\begin{equation*}
  A\in\mathbb{Z}_q^{m\times n}, \qquad s\in\mathbb{Z}_q^n, \qquad e\in\mathbb{R}^m,
\end{equation*}
each secret component $s_i$ is represented using $\lceil \log_2 q\rceil$ binary variables, while each noise component $e_j$ is encoded using $\lceil \log_2 q_e\rceil$ binary variables, where $q_e$ denotes the effective discretized noise range. The total number of logical binary variables in the resulting QUBO is therefore
\begin{equation}
  N_{\text{logical}} = n\lceil \log_2 q\rceil + m\lceil \log_2 q_e\rceil.
\end{equation}
The quadratic objective arises from expanding $\|As+e-b\|^2$, which induces couplings between all binary variables associated with nonzero entries in $A$. Even for moderately dense matrices, this produces a highly connected QUBO whose interaction graph is far from sparse. As a result, the dominant limitation is not only the number of logical variables, but also the cost of embedding this dense logical graph onto physical hardware.

\begin{figure}[t]
  \centering
  \includegraphics[width=0.85\linewidth]{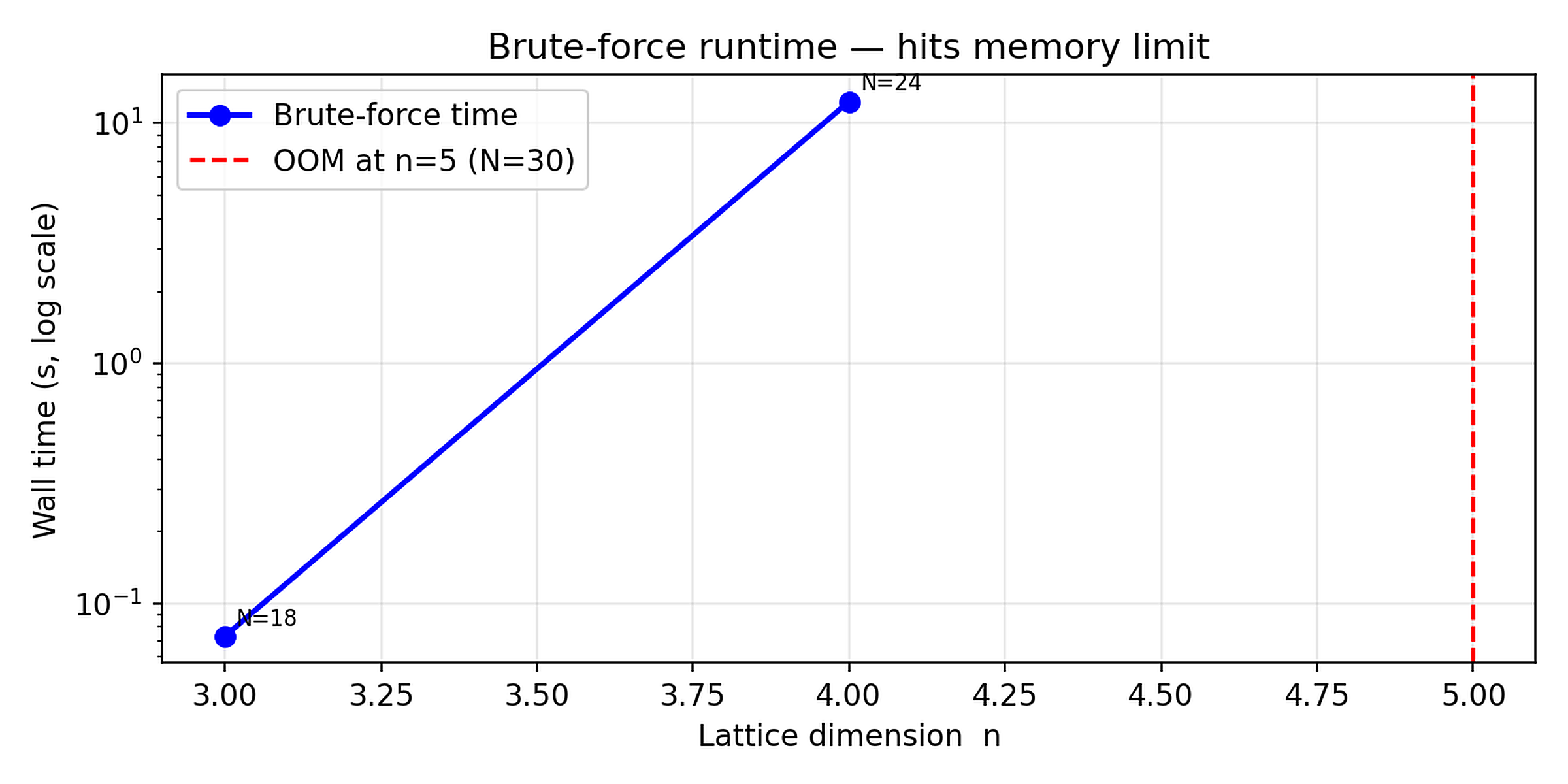}
  \caption{Brute-force runtime as a function of lattice dimension in the exact-enumeration setting. The wall-clock time grows sharply, and the study reaches an out-of-memory boundary around $n=5$ ($N=30$ logical variables).}
  \label{fig:scalingwall}
\end{figure}

Although exhaustive optimization provides an exact benchmark for validating the encoding, its computational complexity increases exponentially with the number of binary variables. In the toy scaling experiment shown here, each dimension contributes 2 secret bits and 4 noise bits, so $N=6n$ when $n=m$ and $q=4$. Exact search therefore remains practical only for relatively small instances, and the computational cost becomes prohibitive around $N\approx 30$ binary variables. Figure~\ref{fig:scalingwall} illustrates this growth in brute-force runtime. This exponential scaling motivates the use of heuristic optimization algorithms or dedicated quantum optimization hardware for larger problem instances.

\begin{figure}[t]
  \centering
  \includegraphics[width=\linewidth]{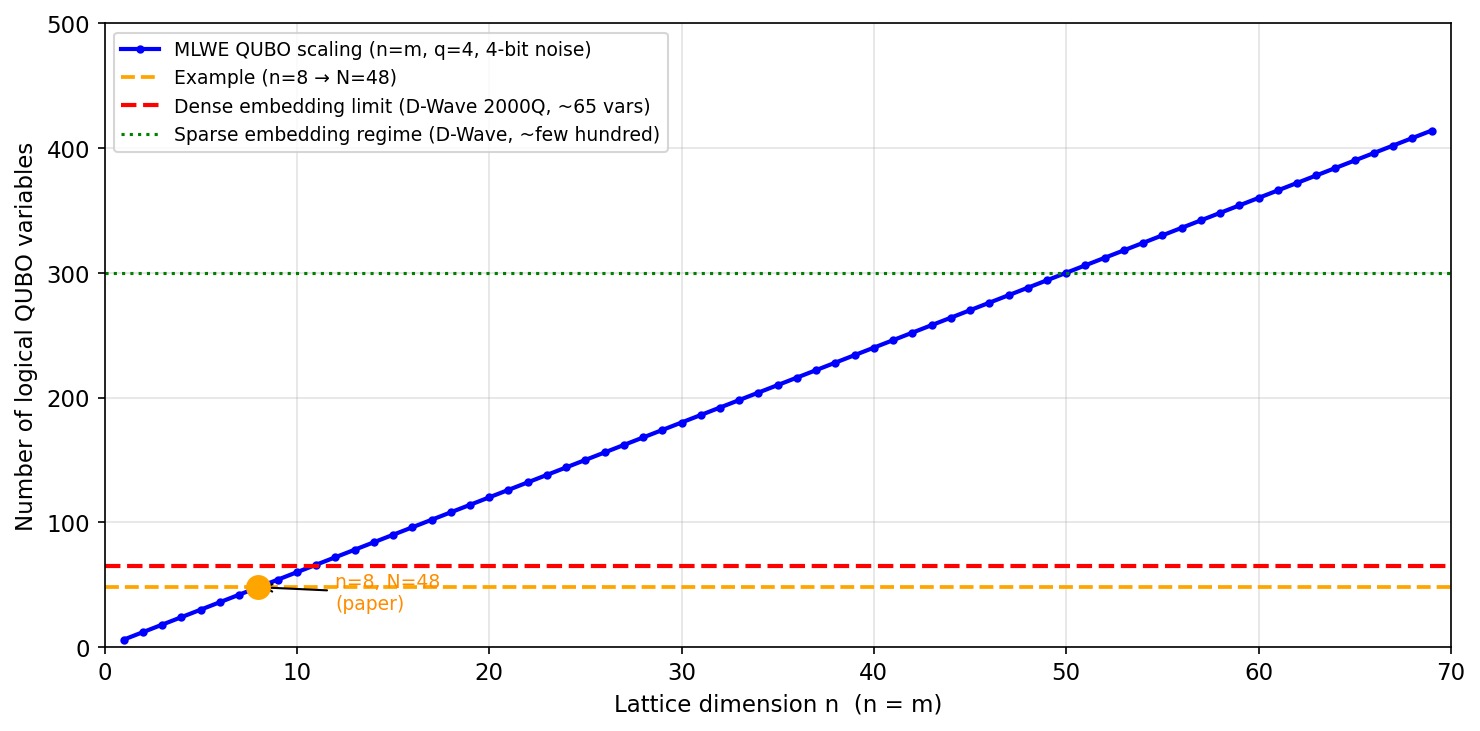}
  \caption{Growth of logical variable count $N_{\text{logical}}$ as a function of MLWE dimension $n$ obtained from exact simulation. The red dashed line indicates the approximate dense-QUBO embedding regime associated with current quantum annealing architectures. While $N_{\text{logical}}$ scales linearly with $n$ for fixed modulus $q$ and fixed noise precision, practical embedding requirements increase due to connectivity and encoding overhead.}
  \label{fig:scalability}
\end{figure}

The practical applicability of the formulation is further constrained by the hardware resources required to embed dense QUBO problems. Figure~\ref{fig:scalability} compares the growth in logical variables required by the proposed encoding with the approximate dense-connectivity limits of current quantum annealing architectures. Although the logical-variable count increases approximately linearly with the lattice dimension for fixed modulus and noise precision, the dense connectivity of the resulting QUBO causes hardware limitations to be reached much earlier than suggested by the logical-variable count alone. Consequently, embedding overhead becomes the dominant practical constraint for realistic parameter sizes.

Overall, the exact simulations validate the correctness of the proposed MLWE-to-QUBO formulation for representative noiseless and noisy instances. In both cases, exhaustive optimization successfully identifies the prescribed secret as the unique global minimum of the QUBO objective, confirming the consistency of the encoding. The computational scaling analysis further shows that exhaustive enumeration rapidly becomes infeasible as the number of binary variables increases, while the hardware scalability study demonstrates that dense QUBO embeddings remain limited by current annealing architectures. These observations indicate that the proposed formulation is best viewed as a rigorous framework for analyzing MLWE optimization landscapes and for benchmarking future quantum and hybrid optimization methods, rather than as a direct approach for solving cryptographic-scale parameter sets with existing hardware.

\section{Discussion and Conclusion}
Our theoretical analysis and numerical simulations support three main conclusions. First, the proposed QUBO model exactly captures the planted secret for the small noiseless and noisy benchmark instances considered here, including the explicit recovery of discretized error variables. Second, robustness can be described in two equivalent ways: geometrically through the convex stability polytope in noise space, and energetically through the gap between the best and second-best QUBO solutions. Third, the formulation scales poorly for dense instances: logical-variable growth is linear in dimension for fixed encoding precision, but exhaustive search and hardware embedding requirements become prohibitive quickly.

Overall, the contribution is best viewed as a structured MLWE-to-QUBO methodology rather than a practical cryptanalytic threat to deployed post-quantum systems. The framework clarifies how secret and noise variables couple inside the optimization landscape, provides interpretable stability diagnostics, and highlights the gap between elegant small-scale encodings and current hardware feasibility. Future work could target sparse or structured QUBOs, Ring-MLWE variants, and hybrid classical--quantum heuristics that preserve the interpret ability of the present construction.

\appendix
\section{Compact Derivation and Hardware Estimate}
For the noisy $3\times 3$ benchmark, we optimize
\begin{equation*}
  E(s,e)=\|As+e-b\|^2 + \lambda\|e\|^2,
\end{equation*}
with 2-bit encoding for each secret coefficient and 4-bit discretization for each noise entry. This yields a joint binary vector of length 18 and a block-structured QUBO of the form
\begin{equation*}
  \mathcal{Q} = \begin{bmatrix}\mathcal{Q}_{ss} & \mathcal{Q}_{se} \\ \mathcal{Q}_{es} & \mathcal{Q}_{ee}\end{bmatrix},
\end{equation*}
where $\mathcal{Q}_{ss}$ encodes secret interactions, $\mathcal{Q}_{ee}$ encodes discretized noise and regularization, and $\mathcal{Q}_{se}$ captures the residual coupling. For the benchmark instance, numerical minimization returns the planted solution $s=(1,0,1)$ together with the discretized perturbation $e=(0.2,-0.2,0.3)$.

For larger systems, the logical-variable count obeys
\begin{equation*}
  N_{\text{logical}} = n\lceil\log_2 q\rceil + m\lceil\log_2 q_e\rceil.
\end{equation*}
Using a simple large-parameter illustrative estimate with $n=256$, $m=512$, $q=3329$, and 4 noise bits, one obtains
\begin{equation*}
  N = 256\cdot 12 + 512\cdot 4 = 5120.
\end{equation*}
This back-of-the-envelope count should be read only as an indicative scaling calculation rather than as a precise module-level parameterization of Kyber. Even at this conservative scale, the required logical problem size remains far beyond the dense logical capacity of current annealing hardware, underscoring why the present formulation is best viewed as an analysis tool rather than as a practical route to cryptographic-scale attacks.

\section{Explicit Worked Example for the Noisy Benchmark}
For completeness, we also record the explicit noisy example underlying the QUBO construction, following the original PDF. Consider
\begin{equation*}
A=
\begin{bmatrix}
1 & 2 & 0 \\
0 & 1 & 1 \\
2 & 0 & 1
\end{bmatrix}, \qquad
s=
\begin{bmatrix}
1 \\ 0 \\ 1
\end{bmatrix}, \qquad
e=
\begin{bmatrix}
0.2 \\ -0.2 \\ 0.3
\end{bmatrix}.
\end{equation*}
Then
\begin{equation*}
  b = As + e =
  \begin{bmatrix}
  1.2 \\ 0.8 \\ 3.3
  \end{bmatrix}.
\end{equation*}
The joint objective is
\begin{equation*}
  E(s,e)=\|As+e-b\|^2 + \lambda\|e\|^2,
\end{equation*}
with $\lambda=0.1$, chosen here as a mild regularization for the toy benchmark. We encode the secret using 2-bit coefficients,
\begin{equation*}
  s_1=x_1+2x_2, \qquad s_2=x_3+2x_4, \qquad s_3=x_5+2x_6,
\end{equation*}
and the noise with 4-bit shifted discretizations,
\begin{align*}
  e_1 &= 0.1(y_1+2y_2+4y_3+8y_4)-0.7, \\
  e_2 &= 0.1(y_5+2y_6+4y_7+8y_8)-0.7, \\
  e_3 &= 0.1(y_9+2y_{10}+4y_{11}+8y_{12})-0.7,
\end{align*}
so that each $e_i$ ranges over $\{-0.7,-0.6,\ldots,0.8\}$. The residual components are
\begin{align*}
  r_1 &= (s_1+2s_2)+e_1-1.2, \\
  r_2 &= (s_2+s_3)+e_2-0.8, \\
  r_3 &= (2s_1+s_3)+e_3-3.3,
\end{align*}
and the total energy becomes
\begin{equation*}
  E(s,e)=r_1^2+r_2^2+r_3^2+\lambda(e_1^2+e_2^2+e_3^2).
\end{equation*}
After substitution of the binary encodings, this yields an 18-variable QUBO in the combined vector
\begin{equation*}
  z=[x_1,\ldots,x_6,y_1,\ldots,y_{12}]^\top,
\end{equation*}
with the same block form
\begin{equation*}
  \mathcal{Q} = \begin{bmatrix}\mathcal{Q}_{ss} & \mathcal{Q}_{se} \\ \mathcal{Q}_{es} & \mathcal{Q}_{ee}\end{bmatrix}.
\end{equation*}
The minimizing binary strings decode to
\begin{equation*}
  x=[1,0,0,0,1,0], \qquad y=[1,0,0,1,1,0,1,0,0,1,0,1],
\end{equation*}
and therefore recover the planted instance
\begin{equation*}
  s=(1,0,1), \qquad e=(0.2,-0.2,0.3).
\end{equation*}
This explicit benchmark is the concrete reference example used throughout the manuscript to illustrate the joint secret--noise QUBO construction. In particular, the 4-bit shifted encoding maps the recovered $y$ blocks to the intended signed fractional noise values in steps of $0.1$.


\begin{thebibliography}{99}

\bibitem{Kadowaki1998}
T.~Kadowaki and H.~Nishimori,
``Quantum annealing in the transverse Ising model,''
\emph{Physical Review E}, vol.~58, no.~5, pp.~5355--5363, 1998.

\bibitem{Albash2018}
T.~Albash and D.~A.~Lidar,
``Adiabatic quantum computation,''
\emph{Reviews of Modern Physics}, vol.~90, no.~1, p.~015002, 2018.

\bibitem{Venturelli2015}
D.~Venturelli, S.~Mandra, S.~Knysh \emph{et al.},
``Quantum optimization of fully connected spin glasses,''
\emph{Physical Review X}, vol.~5, no.~3, p.~031040, 2015.

\bibitem{PerdomoOrtiz2012}
A.~Perdomo-Ortiz, N.~Dickson, M.~Drew-Brook, G.~Rose, and A.~Aspuru-Guzik,
``Construction of energy functions for lattice heteropolymer models: A case study in quantum annealing,''
\emph{Scientific Reports}, vol.~2, p.~571, 2012.

\bibitem{Ding2024}
J.~Ding, G.~Spallitta, and R.~Sebastiani,
``Effective prime factorization via quantum annealing by modular locally-structured embedding,''
\emph{Scientific Reports}, vol.~14, p.~3518, 2024.

\bibitem{Shor1994}
P.~W.~Shor,
``Algorithms for quantum computation: discrete logarithms and factoring,''
in \emph{Proceedings of the 35th Annual Symposium on Foundations of Computer Science (FOCS)}, 1994, pp.~124--134.

\bibitem{Regev2005}
O.~Regev,
``On lattices, learning with errors, random linear codes, and cryptography,''
in \emph{Proceedings of the 37th Annual ACM Symposium on Theory of Computing (STOC)}, 2005, pp.~84--93.

\bibitem{Regev2009}
O.~Regev,
``Lattices in computer science,''
\emph{Bulletin of the American Mathematical Society}, vol.~46, no.~1, pp.~1--40, 2009.

\bibitem{Lyubashevsky2010}
V.~Lyubashevsky, C.~Peikert, and O.~Regev,
``On ideal lattices and learning with errors over rings,''
in \emph{EUROCRYPT 2010}. Springer, 2010, pp.~1--23.

\bibitem{Peikert2016}
C.~Peikert,
``A decade of lattice cryptography,''
\emph{Foundations and Trends in Theoretical Computer Science}, vol.~10, no.~4, pp.~283--424, 2016.

\bibitem{Bos2018}
J.~W.~Bos, L.~Ducas, E.~Kiltz \emph{et al.},
``CRYSTALS--Kyber: A CCA-secure module-lattice-based KEM,''
in \emph{IEEE European Symposium on Security and Privacy (EuroS\&P)}, 2018, pp.~353--367.

\bibitem{Lenstra1982}
A.~K.~Lenstra, H.~W.~Lenstra, and L.~Lov{\'a}sz,
``Factoring polynomials with rational coefficients,''
\emph{Mathematische Annalen}, vol.~261, no.~4, pp.~515--534, 1982.

\bibitem{Schnorr2001}
C.-P.~Schnorr,
``Analysis of lattice reduction algorithms,''
in \emph{Mathematical Foundations of Computer Science (MFCS)}. Springer, 2001, pp.~1--20.

\bibitem{Albrecht2015}
M.~R.~Albrecht, R.~Player, and S.~Scott,
``On the concrete hardness of learning with errors,''
\emph{Journal of Mathematical Cryptology}, vol.~9, no.~3, pp.~169--203, 2015.

\bibitem{Blum2003}
A.~Blum, A.~Kalai, and H.~Wasserman,
``Noise-tolerant learning, the parity problem, and the statistical query model,''
\emph{Journal of the ACM}, vol.~50, no.~4, pp.~506--519, 2003.

\bibitem{Laarhoven2015}
T.~Laarhoven,
``Search problems in cryptography: from classical to quantum,''
\emph{Cryptology ePrint Archive}, no.~2015/212, 2015.

\bibitem{Childs2003}
A.~M.~Childs, E.~Farhi, and S.~Gutmann,
``Exponential algorithmic speedup by quantum walk,''
in \emph{Proceedings of the 35th Annual ACM Symposium on Theory of Computing (STOC)}, 2003, pp.~59--68.

\bibitem{Lucas2014}
A.~Lucas,
``Ising formulations of many NP problems,''
\emph{Frontiers in Physics}, vol.~2, p.~5, 2014.

\bibitem{Joseph2021}
D.~Joseph, A.~Callison, C.~Ling, and F.~Mintert,
``Two quantum Ising algorithms for the shortest-vector problem,''
\emph{Physical Review A}, vol.~103, p.~032433, 2021.

\bibitem{Zheng2024}
M.~Zheng, J.~Zeng, W.~Yang, P.-J.~Chang, Q.~Lu, B.~Yan, H.~Zhang, M.~Wang, S.~Wei, and G.-L.~Long,
``Quantum-classical hybrid algorithm for solving the learning-with-errors problem on NISQ devices,''
\emph{Communications Physics}, vol.~8, Art.~208, 2025. Available: \url{https://arxiv.org/abs/2408.07936}

\bibitem{Jiang2026}
S.~Jiang,
``When the learning with errors problem meets the coherent Ising machine: A penalty-free algorithm-hardware co-design,''
\emph{arXiv preprint arXiv:2606.22843}, 2026. Available: \url{https://arxiv.org/abs/2606.22843}

\bibitem{Qayyum2025}
A.~Qayyum,
``Advancing LWE cryptanalysis: An updated MIP model and QUBO formulation for quantum annealing,''
\emph{Prikladnaya Diskretnaya Matematika}, no.~18, pp.~194--200, 2025.

\end{thebibliography}
\end{document}